\documentclass[epj]{webofc}
\usepackage[varg]{txfonts}   

\providecommand{\physrep}{Phys.\ Rep.}

\providecommand{\prc}{Phys.\ Rev.\ C}

\wocname{OMEG 2015}
\woctitle{OMEG 2015}

\usepackage{array}
\usepackage{float}
\usepackage{amsmath}
\usepackage{graphicx}
\usepackage{amssymb}
\usepackage{subfigure}
\usepackage{sidecap}
\usepackage[english]{babel}
\usepackage{txfonts}
\usepackage{calc}
\usepackage[colorlinks]{hyperref}

\begin{document}

\title{Expected impact from weak reactions with light nuclei in core-collapse supernova simulations}

\author{
T. Fischer\inst{1}\fnsep\thanks{\email{fischer@ift.uni.wroc.pl}}
\and
G. Mart{\'i}nez-Pinedo\inst{2,3}
\and
M. Hempel\inst{4}
\and
L. Huther\inst{2}
\and
G. R{\"o}pke\inst{5}
\and
S. Typel\inst{3}
\and
A. Lohs\inst{4}
}

\institute{
Institute for Theoretical Physics, University of Wroc{\l}aw, plac Maksa Borna 9, 50-204 Wroc{\l}aw, Poland
\and
Technische Universit{\"a}t Darmstadt, Schlossgartenstra{\ss}e 2, 2, 64289 Darmstadt, Germany
\and
GSI Helmholtzzentrum f\"ur Schwerionenforschung, Planckstra{\ss}e~1, 64291 Darmstadt, Germany
\and
Department of Physics, University of Basel, Klingelbergstrasse 82, 4058 Basel, Switzerland
\and
Institut f{\"u}r Physik, Universit{\"a}t Rostock, Wismarsche Strasse 43-45, 18057 Rostock, Germany
}

\abstract{
We study the role of light nuclear clusters in simulations of core-collapse supernovae. Expressions for the reaction rates are developed for a large selection of charged current absorption and scattering processes with light clusters. Medium modifications are taken into account at the mean-field level. We explore the possible impact on the supernova dynamics and the neutrino signal during the mass accretion phase prior to the possible explosion onset as well as during the subsequent protoneutron star deleptnoization after the explosion onset has been launched.
}

\maketitle

\section{Introduction}
Massive stars end their lives in the event of a core collapse supernova, triggered from the gravitational collapse of the stellar core. The possible explosion is associated with energy transfer from the hot and lepton rich protoneutron star (PNS), which forms at the very core, into the layer above the PNS surface. It leads to the ejection of the stellar mantle~\cite{Janka:2007,Janka:2012}. Once the explosion proceeds the nascent PNS starts to deleptonize via the emission of neutrinos of all flavors over a timescale of 10--30~seconds~\cite{Pons:1998mm}. From the next Galactic event, most of the neutrinos will be detected during this phase. It is therefore of paramount interest to predict reliable neutrino spectra and luminosities. This phase is also subject of the formation of heavy elements beyond iron in the neutrino-driven wind, a low mass outflow ejected via neutrino heating from the PNS surface during deleptonization~\cite{Woosley:1994ux,Arcones:2006uq,MartinezPinedo:2012}. It is necessary to include sophisticated neutrino transport in the models, based on which it became possible only recently to simulate the entire PNS deleptonization phase~\cite{Fischer:2009af,Huedepohl:2010,Fischer:2012a}. Therefore the nuclear equation of state (EOS) and weak rates must be treated consistently~\cite{Reddy:1998,MartinezPinedo:2012,Roberts:2012}. It has long been argued about the role of light nuclear clusters, e.g., deuteron ($^2$H), triton ($^3$H), $^3$He and $^4$He. It has been shown that in particular $^2$H and $^3$H can be as abundant as free protons~\cite{Hempel:2012,Fischer:2014}. In a first attempt inelastic (anti)neutrino absorption processes were considered using vacuum cross sections and simplified rate expressions, ignoring detailed balance and final state Pauli blocking~\cite{Furusawa:2013}. The possible impact of weak reactions with light nuclei on the nucleosynthesis relevant conditions in the neutrino-driven wind were studied in Ref.~\cite{Arcones:2008}. Here we extend the selection of weak processes and include medium modifications. We derive expressions for the reaction rates. These are then included in the core-collapse supernova simulations which are discussed in this article. A strong impact from the inclusion of weak rates with light clusters cannot be found.
\section{Core-collapse supernova model}
Our core-collapse supernova model is AGILE-BOLTZTRAN. It is based on spherically symmetric general relativistic neutrino radiation hydrodynamics with three flavor Boltzmann neutrino transport~\cite{Liebendoerfer:2001a,Liebendoerfer:2001b,Liebendoerfer:2002,Liebendoerfer:2004,Liebendoerfer:2005a}. The list of weak processes considered is given in Table~\ref{tab:weak-reactions}.
\begin{SCtable}
\small
\begin{tabular}[t]{lc}
\hline
\hline
\small (1)  $\nu_e n \rightleftarrows e^- p$ & \cite{Reddy:1998} \\ 
\small (2)  $\bar\nu_e p \rightleftarrows e^+ n$ & \cite{Reddy:1998} \\
\small (3)  $\nu_e  (A,Z-1) \rightleftarrows e^- (A,Z)$ & \cite{Juodagalvis:2010} \\
\small (4)  $\nu N \rightleftarrows \nu' N$ & \cite{Bruenn:1985en,Mezzacappa:1993gm} \\
\small (5)  $\nu (A,Z) \rightleftarrows \nu' (A,Z)$ & \cite{Bruenn:1985en,Mezzacappa:1993gm} \\
\small (6)  $\nu\;^4\text{He} \rightleftarrows \nu' \;^4\text{He}$ & \cite{Bruenn:1985en,Mezzacappa:1993gm} \\
\small (7)  $\nu e^\pm \rightleftarrows \nu' e^\pm$ & \cite{Bruenn:1985en,Mezzacappa:1993gx} \\
\small (8)  $\nu \bar{\nu} \rightleftarrows e^- e^+ $ & \cite{Bruenn:1985en} \\
\small (9)  $\nu \bar{\nu} N N  \rightleftarrows  N N$ & \cite{Hannestad:1997gc} \\
\small (10)  $\nu_e \bar\nu_e \rightleftarrows  \nu_{\mu/\tau} + \bar\nu_{\mu/\tau}$ & \cite{Buras:2002wt,Fischer:2009} \\
\small (11) $\nu \bar\nu (A,Z) \rightleftarrows (A,Z)^*$ & \cite{Fuller:1991,Fischer:2013} \\
\hline
\label{tab:nu-reactions}
\end{tabular}
\quad
\begin{tabular}[t]{lc}
\hline
\hline
\small (1) $\nu_e\,^2\text{H}\rightleftarrows p\,p\,e^-$ \\
\small(2) $\bar\nu_e\,^2\text{H}\rightleftarrows n\,n\,e^+$ \\
\small(3) $\nu_e\,n \,n \rightleftarrows \, ^2\text{H}\,e^-$ \\
\small(4) $\bar\nu_e\,p \,p \rightleftarrows \, ^2\text{H}\,e^+$ \\
\small(5) $\nu_e\,^3\text{H}\rightleftarrows n\,p\,p\,e^-$ \\
\small(6) $\bar\nu_e\,^3\text{H}\rightleftarrows n\,n\,n\,e^+$ \\
\small(7) $\nu_e\,^3\text{H}\rightleftarrows \,^3\text{He}\,e^-$ \\
\small(8) $\bar\nu_e\,^3\text{He}\rightleftarrows \,^3\text{H}\,e^+$ \\
\hline
\small(9) $\nu\,\,^2\text{H}\rightleftarrows \, ^2\text{H}\,\nu$ \\
\small(10) $\nu\,\,^3\text{H}\rightleftarrows \, ^3\text{H}\,\nu$ \\
\small(11) $\nu\,\,^3\text{He}\rightleftarrows \, ^3\text{He}\,\nu$ \\
\small(12) $\nu\,\,^2\text{H}\rightleftarrows p\,n\,\nu$ \\
\hline
\end{tabular}
\caption{
{\addtolength{\leftskip}{5mm}\emph{(left panel)} \small Set of standard weak reactions considered in simulations of core-collapse supernovae, including references. ($\nu=\{\nu_e,\bar{\nu}_e,\nu_{\mu/\tau},\bar{\nu}_{\mu/\tau}\}$ unless stated otherwise and $N=\{n,p\}$, and $(A,Z)$ are atomic mass and charge). Note that there are no charged-current absorption reactions with $^4\text{He}$ because of the high binding energy.}
\smallskip
{\addtolength{\leftskip}{5mm}\emph{(right panel)} List of novel weak reactions with light nuclear clusters considered -- charged current absorption (top) and neutral current (bottom).}}
\label{tab:weak-reactions}
\end{SCtable}
For the reaction rates one commonly uses the elastic approximation, e.g., for reaction (1) and (2) in the left panel of Table~\ref{tab:weak-reactions}:
\begin{eqnarray}
1/\lambda_{\nu_en}(E_{\nu_e}) \simeq
\frac{G_F^2}{\pi} \frac{V_{ud}^2}{(\hbar c)^4} (g_V^2+3 g_A^2) \,p_e \, E_e \, \left(1-f_e(E_{e})\right)
\frac{n_n-n_p}{1-\exp\left\{\beta(\mu_p^0 - \mu_n^0)\right\}} \,\,\,,
\end{eqnarray}
where $n_n$ and $n_p$ are the neutron and proton number densities respectively. The function $f_e(E_e)$ is the equilibrium electron Fermi-function, $\left\{1+\exp{\{\beta(E_e-\mu_e)\}}\right\}^{-1}$, with inverse temperature $\beta=1/T$ and electron chemical potentials $\mu_e$. The medium modifications enter via the EOS, i.e. the nucleon dispersion relations and particle densities, or equivalent via the chemical potentials, as follows
\begin{eqnarray}
E_N({\bf p}_N) = \sqrt{{\bf p}^2_N + {m^*_N}^2} + U_N  \;\;\;\;\;,\;\;\;\; \mu_N^0 = \mu_N - U_N - m_N^*~,
\label{eq:E}
\end{eqnarray}
where $\mu_N$ are the full chemical potentials that contain contributions from interactions. Both effective mass $m_N^*$ and mean field potential $U_N$ are determined consistently with the EOS. Since neutrino transport uses reaction rates with respect to the incoming (anti)neutrino energy, $E_{\nu_e}$($E_{\bar\nu_e}$) one can apply energy conservation and use the particle's dispersion relations Eq.~(\ref{eq:E}) to relate them with the electron(positron) energy, $E_{e^-}(E_{e^+})$, as follows,
\begin{eqnarray}
E_{\nu_e} = E_{e^-} - (m_n^* - m_p^*) - \triangle U \;\;\;\;\;,\;\;\;\; E_{\bar{\nu}_e} = E_{e^+} +  (m_n^* - m_p^*) + \triangle U \;\;\;\;\; (\triangle U=U_n-U_p)~.
\end{eqnarray}
For the reverse process, i.e. the neutrino emissivity, detailed balance applies:
\begin{eqnarray}
&& j_{\nu}(E_{\nu}) = \exp\left\{
-\beta\left(E_{\nu}-\mu_{\nu}^\text{eq.}\right)
\right\}\,1/\lambda_{\nu}(E_{\nu})\;,
\end{eqnarray}
with equilibrium neutrino chemical potentials $\mu_{\nu_e}^\text{eq.} = \mu_e - (\mu_n - \mu_p)$ and $\mu_{\bar\nu_e}^\text{eq.}=-\mu_{\nu_e}^\text{eq.}$.

\section{Weak reactions with light nuclei}
{\bf $\nu$-absorption on deuteron:} In the elastic approximation the $\nu$ opacity for the absorption on $^2$H (reactions (1) and (2) in the right panel of Table~\ref{tab:weak-reactions}) is given by the following expression:
\begin{equation}
1/\lambda(E_\nu) =
\frac{g_{^2\text{H}}}{2}
\int
\frac{d^3p_{^2\text{H}}}{(2\pi \hbar c)^3}
dp_e d(\cos\theta)
\left(\frac{d\sigma_{\nu\,^2\text{H}}}{dp_e}(E_\nu^*)\right)
\tilde{f}_{^2\text{H}}(E_{^2\text{H}})
(1-f_e(E_e))
\left(1-f_1(E_1)\right)
\left(1-f_2(E_2)\right)\;,
\label{eq:opacity_d}
\end{equation}
with final state nucleon distributions $f_1$ and $f_2$ as well as $^2$H distribution $\tilde{f}_{^2\text{H}}$ and degeneracy $g_{^2\text{H}}$ and $\theta$ being the angle between relative and center-of-mass momenta of the final state nucleons. For the cross section appearing in Eq.~\eqref{eq:opacity_d} we use the vacuum cross section provided in Ref.~\cite{Nakamura:2001}. However, these are evaluated at a ''shifted'' neutrino energy in order to account for the medium modifications.
\begin{eqnarray}
E_{\nu_e}^*=E_{\nu_e}+(m^*_{^2\text{H}}-m_{^2\text{H}})+U_{^2\text{H}}-2(m_p^*-m_p)-2U_p\;, 
\label{eq:Eve}\\
E_{\bar\nu_e}^*=E_{\bar\nu_e}+(m^*_{^2\text{H}}-m_{^2\text{H}})+U_{^2\text{H}}-2(m_n^*-m_n)-2U_n\;,
\label{eq:Eveb}
\end{eqnarray}
which depend on the nuclear EOS. This assumes that the medium modifications change the energetics of the reaction without modifying the wave functions of the involved states. The deuteron mean-field potential $U_{^2\text{H}}$ is obtained by comparison with the non-interacting gas of deuteron\footnote{The nuclear EOS used here considers Boltzmann statistics for light clusters and in addition $m^*_{^2\text{H}}=m_{^2\text{H}}$} $\phi_{^2\text{H}}$: $U_{^2\text{H}} = \mu_{^2\text{H}} - \phi_{^2\text{H}}$, with deuteron chemical potential $\mu_{^2\text{H}}=\mu_n+\mu_p$. The remaining three integrals are still computationally expensive. We therefore reduce expression~\eqref{eq:opacity_d} neglecting final state nucleon blocking:
\begin{equation}
1/\lambda(E_\nu) = n_{^2\text{H}} \int dp_e \left(\frac{d\sigma_{\nu\,^2\text{H}}}{dp_e}(E_\nu^*)\right) (1-f_e(E_e))~,
\end{equation}
with the deuteron density $n_{^2\text{H}}$. It is valid in the limit of low degeneracy. This is always the case for reactions with $\nu_e$ since protons are never degenerate. However, note that neutrons become (at least partly) degenerate. \\{\bf $e$-captures on deuteron:} For the inverse electron/positron capture on $^2$H (reactions (3) and (4) in the right panel of Table~\ref{tab:weak-reactions}) a similar expression as \eqref{eq:opacity_d} applies, however, with the following replacements:
\begin{equation}
\left(1-f_N\right)\longrightarrow f_N\;\;\;\;\;,\;\;\;\;\; \tilde{f}_{^2\text{H}}\longrightarrow\left(1+\tilde{f}_{^2\text{H}}\right)~,
\end{equation}
and
\begin{equation}
\frac{d\sigma_{e^-\,^2\text{H}}}{d\Omega_{\nu_e} dp_{\nu_e}}(E_e)
\simeq 
\frac{1}{2}
\frac{d\sigma_{\bar\nu_e\,^2\text{H}}}{d\Omega_e dp_e}(E_e)\;,
\;\;\;\;\;
\frac{d\sigma_{e^+\,^2\text{H}}}{d\Omega_{\bar\nu_e} dp_{\bar\nu_e}}(E_e)
\simeq 
\frac{1}{2}
\frac{d\sigma_{\nu_e\,^2\text{H}}}{d\Omega_e dp_e}(E_e)\;,
\end{equation}
for which relativistic electrons/positrons are assumed. For the medium modifications, similar shifts as relations \eqref{eq:Eve} and \eqref{eq:Eveb} are obtained here for electron and positron energies. \\{\bf Reactions with triton:} Here we can have two possible contributions. Break up reactions (5) and (6) in the right panel of Table~\ref{tab:weak-reactions} and reactions (7) and (8) in the right panel of Table~\ref{tab:weak-reactions} connecting the ground states of $^3$H and $^3$He. We find that this second channel by far dominates the neutrino opacity that we describe as,
\begin{eqnarray}
1/\lambda({E_{\nu}}) =
n_{^3\text{H}} \frac{G_F^2\,V_{ud}^2}{\pi(\hbar\,c)^4}\,p_{e}\,E_{e}\,\left(1-f_{e}(E_{e})\right) B(GT)\;,
\end{eqnarray}
with triton number density $n_{^3\text{H}}$ and $B(GT)=5.87$ (known from the triton decay). Here we neglect final state $^3$He blocking and one can relate (anti)neutrino and electron(positron) energies as follows,
\begin{eqnarray}
E_{\nu_e} = E_{e^-} - (m_{^3\text{He}} - m_{^3\text{H}}) - \triangle U~,\;\; E_{\bar{\nu}_e} = E_{e^+} +  (m_{^3\text{He}} - m_{^3\text{H}}) + \triangle U\;\;\text{with}\;\;\triangle U=U_{^3\text{He}}-U_{^3\text{H}}~.
\end{eqnarray}
\begin{figure}
\subfigure[$T=5$~MeV, $\rho=10^{12}$~g~cm$^{-3}$, $Y_e=0.1$.]{
\includegraphics[width=0.4989\columnwidth]{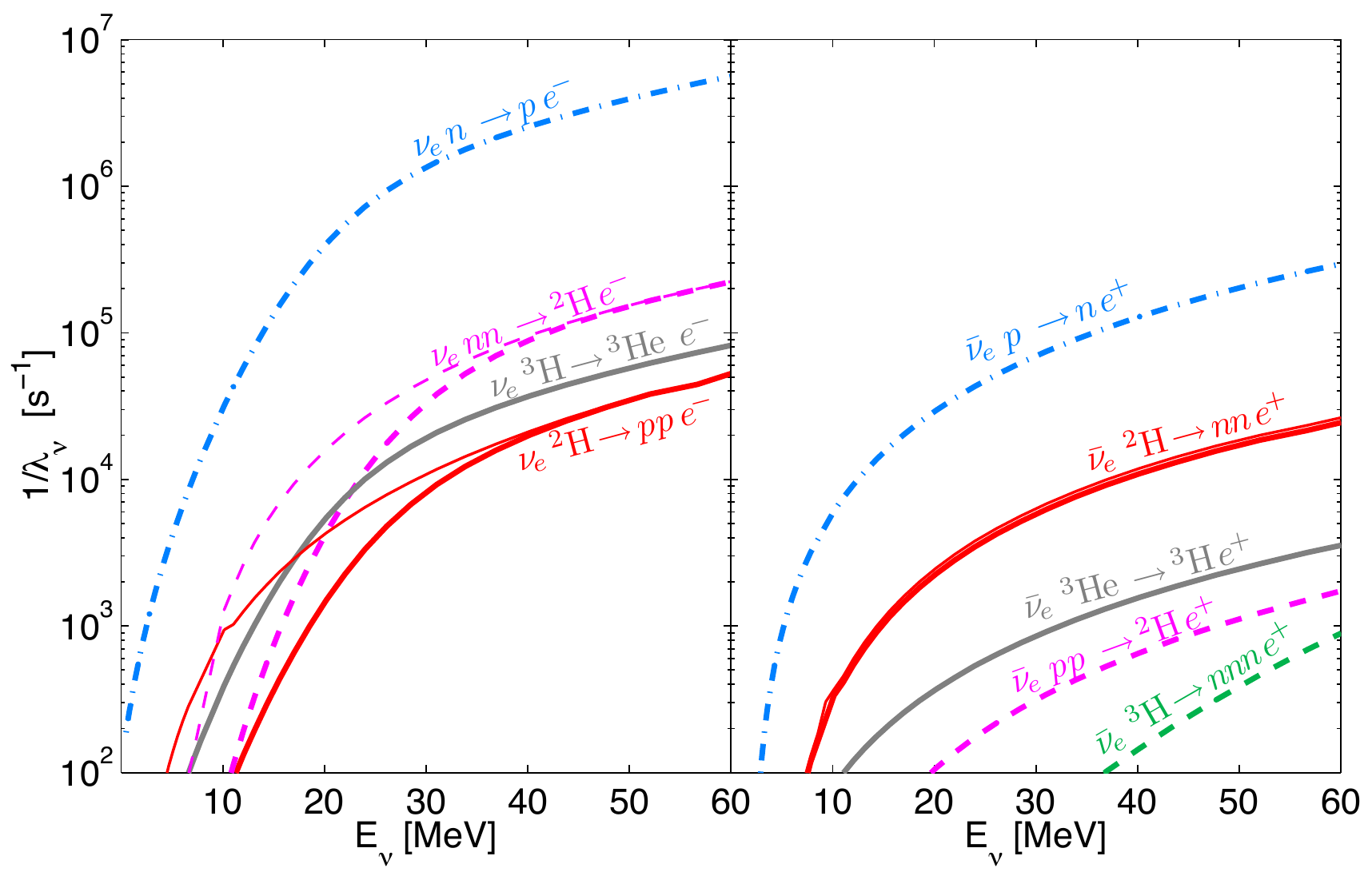}}
\hfill
\subfigure[$T=7$~MeV, $\rho=2\times 10^{13}$~g~cm$^{-3}$, $Y_e=0.075$.]{
\includegraphics[width=0.499\columnwidth]{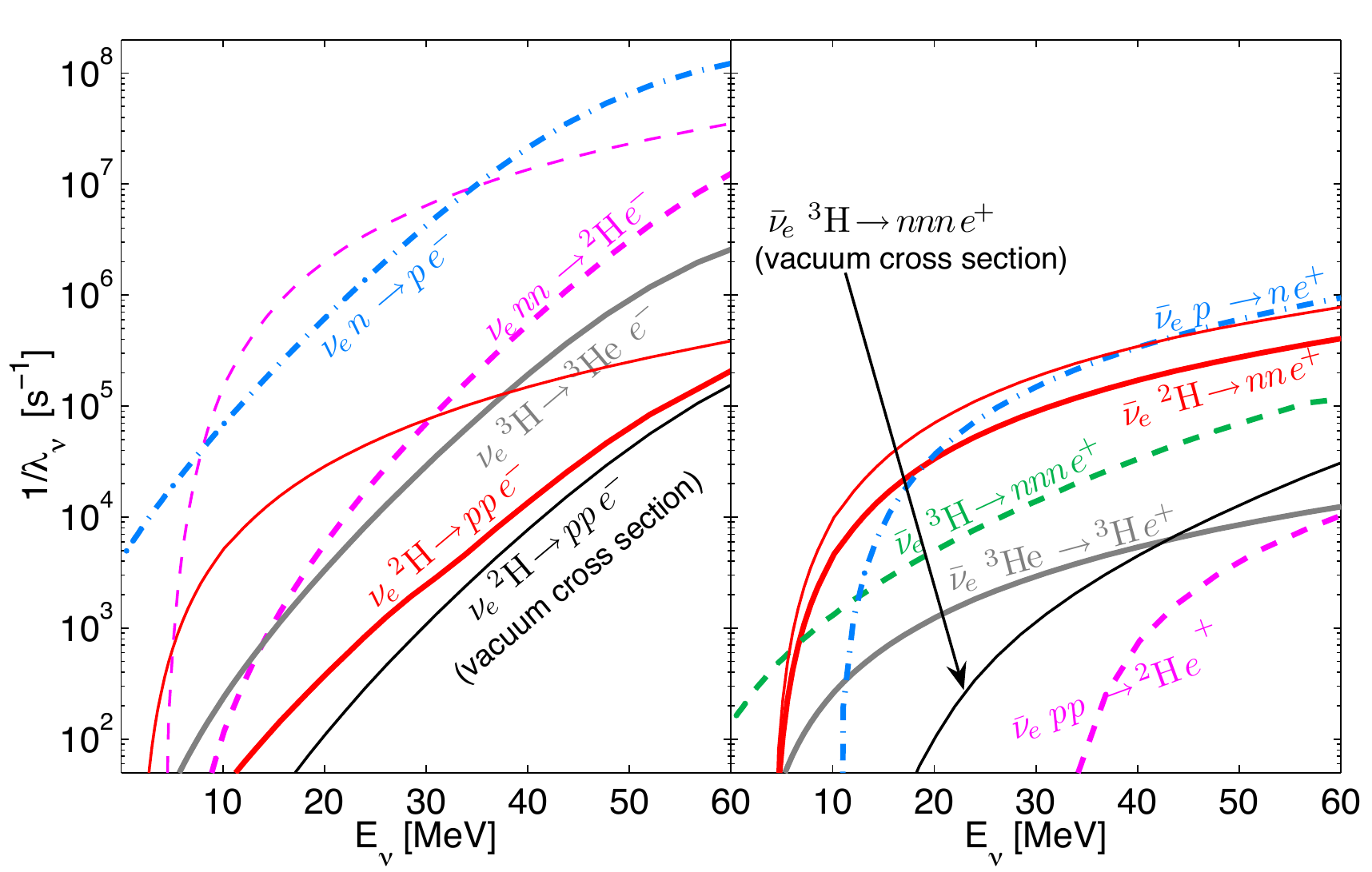}}
\caption{(color online) Charged-current absorption rates for $\nu_e$ (left panels) and $\bar\nu_e$ (right panels) at selected conditions. For $\nu_e$ absorption on deuteron and inverse electron captures on deuteron in the left panels of graphs (a) and (b) we consider the full kinematics (thick red and magenta lines) compared to reduced kinematics without final state electron blocking (thin red and magenta lines). For $\bar\nu_e$ absorption on deuteron in the right panels of graphs (a) and (b) we compare full kinematics (thick red lines) with reduced kinematics without final state nucleon blocking (thin red lines). In addition for $\bar\nu_e$ absorption on triton in the right panel of graph (b) we compare rates with medium modified cross sections (thick green dashed line) and with vacuum cross sections (think black line). The rates for $\nu_e$ absorption on $^3$H with three unbound nucleons in the final state (reaction (5) in the right panel of Table~\ref{tab:weak-reactions}) is not shown here for simplicity. It is much smaller than any other rate and therefore negligible.}
\label{fig:opacity_cc}
\end{figure}

Rates for the charged current absorption reactions are shown in Fig.~\ref{fig:opacity_cc} at two selected conditions. We employ the modified nuclear statistical equilibrium EOS HS(DD2)~\cite{Hempel:2009mc} with the relativistic mean field parametrization DD2~\cite{Typel:2009sy}. We compared this approach with more sophisticated descriptions of the nuclear medium~\cite{Hempel:2011}, i.e. with the generalized relativistic mean field framework~\cite{Typel:2009sy} and with the quantum statistical approach~\cite{Roepke:2009,Roepke:2011}. We find good quantitative agreement, e.g., in terms of the deuteron and triton abundances for conditions relevant for supernova simulations. The opacity for neutral-current neutrino scattering on light clusters (reactions (9)--(11) in the right panel of Table~\ref{tab:weak-reactions}) is considered in this work at the level of the elastic approximation too~\cite{Bruenn:1985en}. In comparison to the neutrino scattering on neutrons, scattering on light clusters has negligible contributions. This is mainly an EOS effect since the abundance of neutrons exceeds those of protons and also light nuclei, which is typical for the neutron-rich supernova conditions. Inelastic contributions due to reaction~(12) in the right pannel of Table~\ref{tab:weak-reactions} are evaluated based on the cross sections from Ref.~\cite{Nakamura:2001}, i.e. the low-energy cut due to the deuteron binding energy threshold and a slight high-energy enhancement of the scattering opacity.

\section{Supernova simulation results}

The simulations are launched from a 11.2~M$_\odot$ progenitor~\cite{Woosley:2002zz}. It was evolved consistently though core collapse, bounce and post bounce phases using the EOS HS(DD2), taking fully into account light nuclear clusters. We compare results from two simulations, i.e. with and without including weak rates with light clusters based on the selection given in the right panel of Table~\ref{tab:weak-reactions}. In general, the post bounce phase of core-collapse supernovae is determined from mass accretion from the still gravitationally unstable layers above the stellar core. The mass of the central PNS grows accordingly. The physics is given in terms of neutrino heating and cooling inside the layer of accumulated low-density material at the PNS surface, i.e. the region of neutrino decoupling. There the abundance of light clusters is small compared to those of free neutrons and protons (c.f. Fig.~12(a) in Ref.~\cite{Fischer:2014}). Only towards higher densities the abundance of $^2$H and $^3$H increases, however, there the neutrinos become trapped. This situation remains during the entire mass accretion phase. It is the reason why we cannot find any impact from the consistent inclusion of weak rates with light nuclei on the supernova dynamics nor on the neutrino signal.
\begin{figure}[t!]
\centering
\includegraphics[width=1\columnwidth]{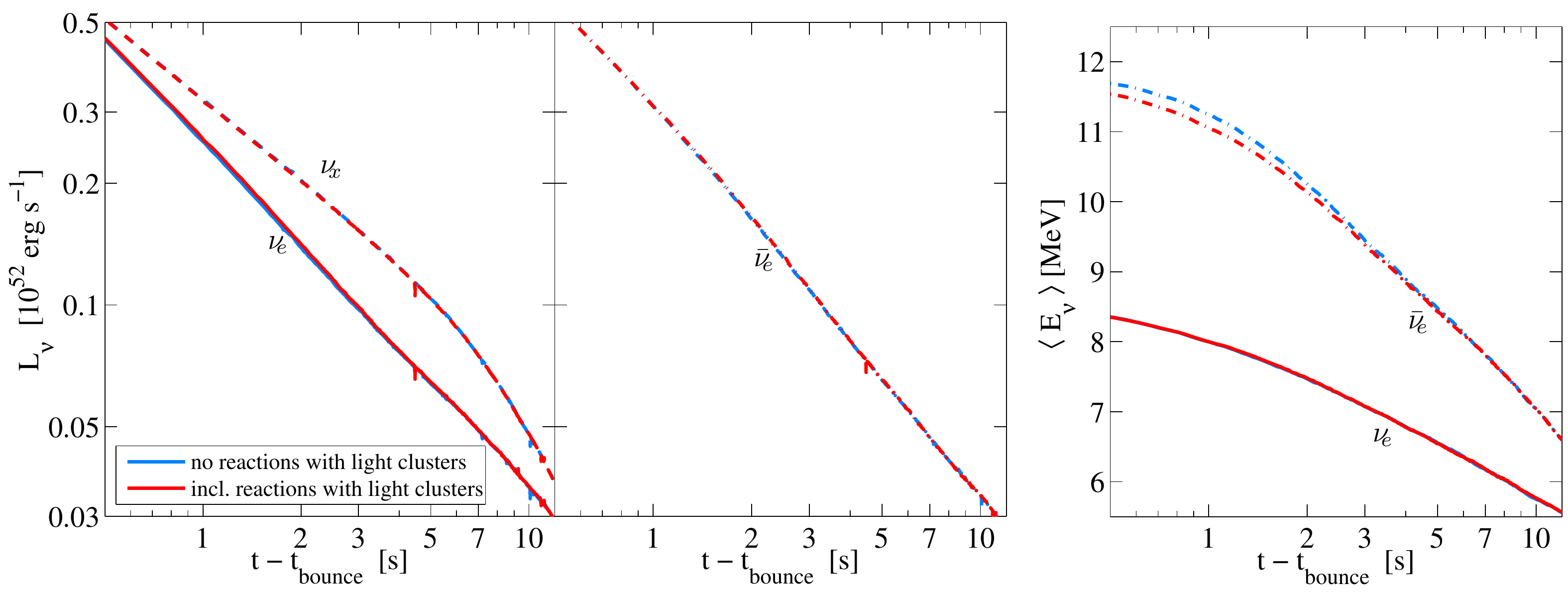}
\caption{(color online) Evolution of neutrino luminosities and energies during the PNS deleptonization phase ($\langle E_{\nu_x} \rangle$ are left out for simplicity, they are not affected from the inclusion of weak rates with light clusters).}
\label{fig:neutrinos}
\end{figure}
The situation changes somewhat during the PNS deleptonization, i.e. after the launch of the supernova explosion, when the neutrinospheres move to generally higher densities. It is associated with the vanishing of mass accretion at the PNS surface. We parametrize the explosion onset via enhanced neutrino heating in the gain region. Once the explosion proceeds we switch back to the standard heating rates. The evolution of neutrino fluxes and energies is illustrated in Fig.~\ref{fig:neutrinos}. The small differences with and without consistent inclusion of weak reactions with light clusters, in particular for the $\bar\nu_e$ channel, is due to inelastic scattering on electrons(positrons) which dominates inelastic $\bar\nu_e$ opacity at all times during the PNS deleptonization phase (see Fig.~\ref{fig:mfp}). For the $\nu_e$-opacity an impact form the inclusion of reactions with light clusters cannot be expected. It is dominated by absorption on neutrons by several orders of magnitude.
\begin{figure}[htp]
\subfigure[$t-t_\text{bounce}=1$~s]{
\includegraphics[width=0.499\columnwidth]{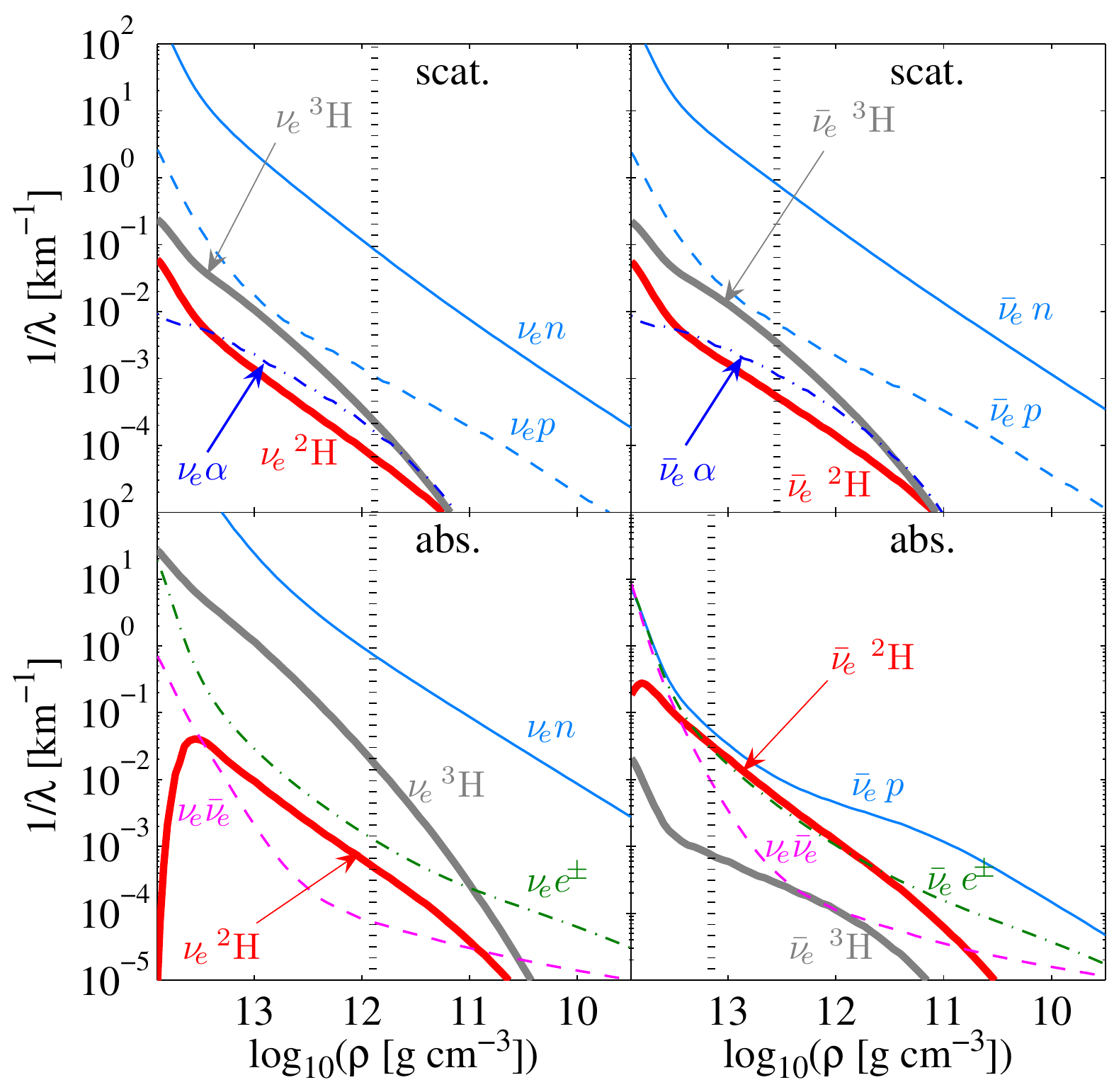}}
\hfill
\subfigure[$t-t_\text{bounce}=5$~s]{
\includegraphics[width=0.499\columnwidth]{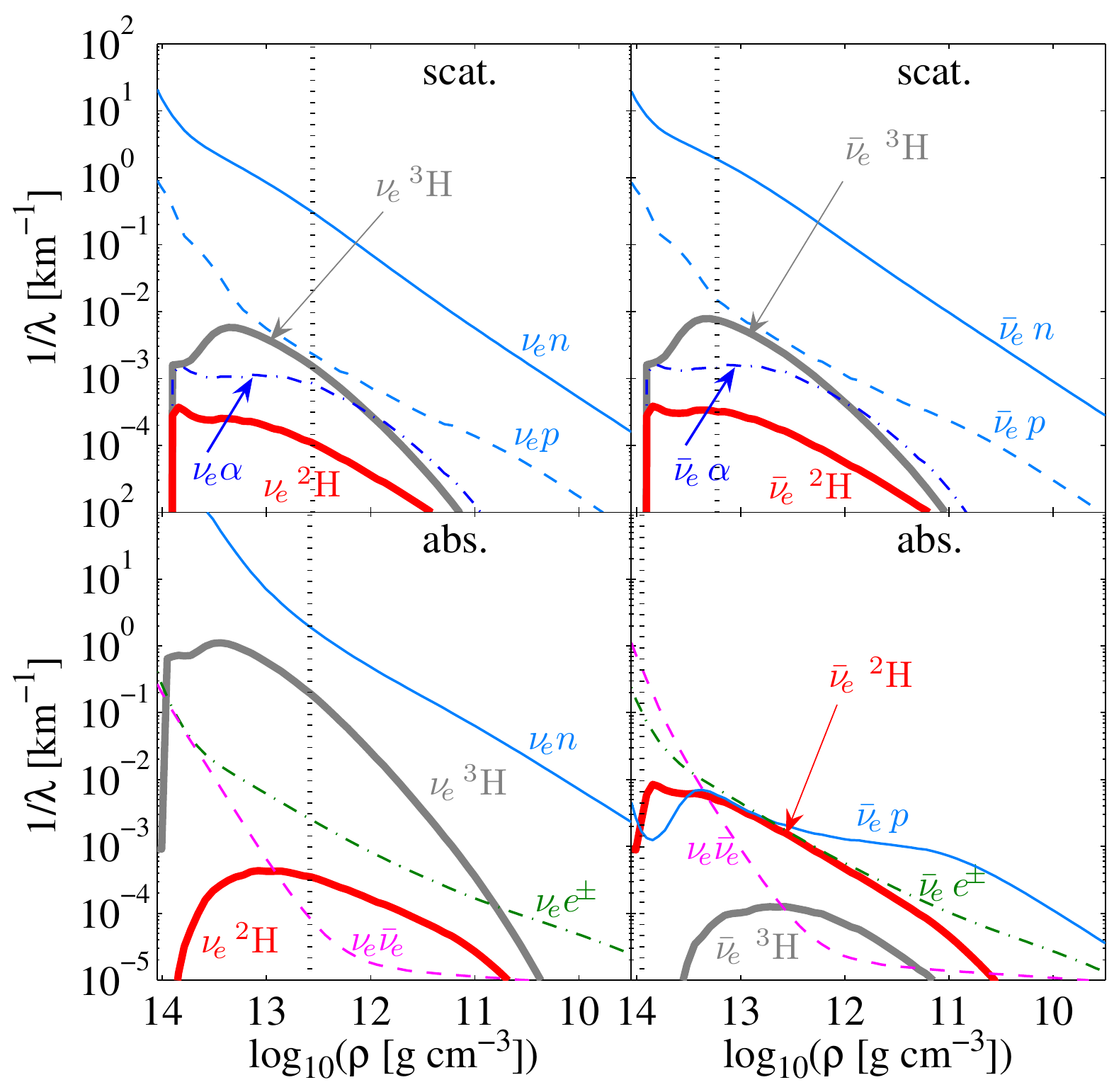}}
\caption{(color online) Density dependence of selected $\nu_e$ (left panels) and $\bar\nu_e$ (right panels) opacities at two selected post bounce times, for elastic scattering reactions (top panels) and charged-current absorption reactions (bottom panels). Vertical dotted lines mark the positions of the averaged neutrinospheres for last inelastic and elastic scattering respectively. Note that ($\nu_e\bar\nu_e$) is the sum of $e^-e^+$-annihilation and $N$--$N$--Bremsstrahlung.}
\label{fig:mfp}
\end{figure}
\section{Summary}
In this article we presented preliminary results from core-collapse supernova simulations that take into account weak reactions with light nuclear clusters. We extended the list of previously considered weak processes with light clusters substantially. Moreover, expressions for the opacity are derived at the level of the elastic approximation, including shifts of the vacuum cross sections due to medium modification. Note that additional processes not considered here, in particular with $^3$He and $^3$He, may be of some relevance. This extends beyond the scope of the present work and will be explored in a more detailed and systematic study.

The overall impact from the inclusion of weak reactions with light clusters remains small, even negligible during the accretion phase of core-collapse supernovae prior to the possible explosion onset. Even though it was explored here in spherical symmetry multidimensional phenomena which may lead to composition mixing are unlikely to change this overall picture. Due to the low abundance of light clusters the associated weak processes cannot compete with the standard neutrino heating and cooling channels. During the later PNS deleptonization, i.e. after the explosion onset, we find a mild reduction of the average $\bar\nu_e$ energy during the early phase. This is associated with a slight enhancement of the charged current absorption opacity. However, the overall charged current $\bar\nu_e$ opacity is dominated by inelastic scattering on electrons/positrons. The $\nu_e$ opacity is not affected at all, being dominated by charged current absorption on neutrons.

\subsection*{Acknowledgement}

The authors are grateful for support by the Nuclear Astrophysics Virtual Institute (NAVI) of the Helmholtz Association. TF acknowledges support from the Polish National Science Center (NCN) under grant number UMO-2013/11/D/ST2/02645.


\end{document}